\documentclass[aps,prd,10pt,showpacs,amsmath,showkeys,twocolumn,floatfix,amssymb, preprintnumbers, nofootinbib, superscriptaddress]{revtex4-1} 
\usepackage{epsfig,dcolumn}
\usepackage{graphicx}
\usepackage{comment} 
\usepackage[usenames]{color}
\usepackage{graphicx}
\usepackage{bm}
 \usepackage{ifpdf}
  \usepackage{floatrow}
\usepackage{makecell}
 \usepackage{caption}

\usepackage[normalem]{ulem}
\usepackage[dvipsnames]{xcolor}
\usepackage[utf8]{inputenc}
\usepackage{hyperref}
\hypersetup{ 
  pdfnewwindow=true,      
  colorlinks=true,        
  linkcolor=PineGreen,    
  citecolor=PineGreen,    
  filecolor=PineGreen,    
  urlcolor=PineGreen      
}

\newcommand{\be}{\begin{eqnarray}}
\newcommand{\ee}{\end{eqnarray}}

 \usepackage{epsfig,dcolumn}
\usepackage{graphicx}
\usepackage{comment} 
\usepackage[usenames]{color}
\usepackage{bm}
\usepackage{ifpdf}
\usepackage{floatrow}
\usepackage{makecell}
\usepackage{caption}
 
\begin{document}

\title{Myth of  scattering   in finite volume}

\author{Peng~Guo}
\email{pguo@csub.edu}

\affiliation{Department of Physics and Engineering,  California State University, Bakersfield, CA 93311, USA}
\affiliation{Kavli Institute for Theoretical Physics, University of California, Santa Barbara, CA 93106, USA}

\date{\today}

\begin{abstract} 
In this notes, we  illustrate why the infinite volume scattering amplitude is in fact dispensable when it comes to formulating few-body quantization condition in finite volume. Only subprocess interactions or   interactions associated subprocess amplitudes are essential and fundamental ingredients of quantization conditions. After these ingredients are determined, infinite volume scattering amplitude can be computed separately. The underlying reasons are rooted in   facts that (1) the final physical process is generated by all subprocess or interactions among particles; (2) the ultimate goal of quantization condition in finite volume is to find stationary solutions of few-body system.  That is to say, in the end, it  all comes down to the  solving of   eigenvalue   problem, $\hat{H} | n \rangle = E_n | n \rangle $.
\end{abstract}

\maketitle


\section{Introduction}\label{intro}  
 In past few years, much efforts \cite{Kreuzer:2008bi,Kreuzer:2009jp,Kreuzer:2012sr,Polejaeva:2012ut,Briceno:2012rv,Hansen:2014eka,Hansen:2015zga,Hansen:2016fzj,Briceno:2017tce,Hammer:2017uqm,Hammer:2017kms,Meissner:2014dea,Mai:2017bge,Mai:2018djl, Doring:2018xxx, Romero-Lopez:2018rcb,Guo:2016fgl,Guo:2017ism,Guo:2017crd,Guo:2018xbv,Blanton:2019igq,Romero-Lopez:2019qrt,Blanton:2019vdk,Mai:2019fba,Guo:2018ibd,Guo:2019hih,Guo:2019ogp,Guo:2020wbl,Guo:2020kph}  have been put into the study of few-hadron scattering in finite volume, aiming to mapping out few-hadron scattering information from lattice QCD results. Such a program is   motivated by the fact that lattice QCD computation is normally  performed  in  a periodic box in Euclidean space,  the  physical amplitude is usually not directly accessed  and   discrete energy spectrum is the primary observables from lattice computation. All the dynamical information is encoded in the discrete energy spectrum. Since lattice QCD is considered as ab-initio calculation of QCD, mapping out  few-hadron dynamical information from lattice energy spectrum may be useful for number of reasons, such as, helping to understand the nature of some resonances   and  determine the fundamental parameters of QCD or QCD inspired effective theory.

 Many good progresses have been made in past few years toward this direction, a couple of things have become quite clear:

(1) Such an effort can be accomplished, however, it is normally done not in a direct way.  In another word, except  in two-body case, few-body scattering amplitude is in fact not  directly computed  from lattice results. Only ingredients are extracted directly,  such as interaction potentials, two-body scattering amplitudes, or $K$-matrix components,  depending on a specific approach.  The scattering or decay amplitudes have to be computed in a separate step by assembling all these ingredients together. Metaphorically speaking, lattice QCD is like   IKEA store in modern days,   only   components of a furniture are offered, and   assembling has to be carried out separately in order to see the whole picture. One  of many reasons for this situation is that unlike two-body scattering amplitude that may be parameterized by  a set of parameters, such as phase shifts,  a simple analytic form of parametrization of few-body scattering amplitudes  is usually not available.  Hence,    scattering amplitudes or decay amplitudes conventionally have to be computed though a set of coupled integral equations with few fundamental ingredients of interactions as kernels. One typical example is Faddeev equations approach \cite{9780706505740,Faddeev:1960su,Newton:1982qc} by using two-body scattering amplitudes as central ingredients which may be parameterized in a relative easier way.  This   can also be understood  by the fact that all complex physical processes are generated by subprocesses, in the case of few-body scattering, few-body scattering amplitudes are connected to subprocess by   integral equations.

(2)    Essentially, all different groups are more or less doing the same thing with different twist. Two steps procedures are adopted.  Step one, formulating quantization condition and extracting central ingredients by fitting lattice data. Step two, computing infinite volume scattering amplitude. The choice of central ingredients,  such as,  interaction parameters of effective theory, parameters of $K$-matrix, or interaction potentials, etc, and how quantization conditions are formulated differ from group to group. Ultimately, these central ingredients must be modeled one way or another in order to fit lattice data.

(3) When it comes to the quantization condition in finite volume, in spite of how  one choose to formulate quantization condition, in the end, it is all about finding stationary solutions of few-body system. Stationary solutions can be obtained equivalently  by either finding pole positions of scattering amplitude or solving homogeneous Faddeev type equations directly. To put it simply, it is about solving eigenvalue problem, $\hat{H} | n \rangle = E_n  | n \rangle$ with periodic boundary condition constraint. This is in fact reflected  by how the energy spectrum are extracted from Euclidean space correlation function in lattice computation,
\begin{align}
\langle \mathcal{O} (t)   \mathcal{O}^\dag (0)   \rangle &  \propto \sum_n  \langle \mathcal{O} (0) e^{- \hat{H} t}   | n \rangle \langle n |   \mathcal{O}^\dag (0)  \rangle   \nonumber \\
& = \sum_n |  \langle \mathcal{O} (0)   | n \rangle  |^2 e^{- E_n t}. 
\end{align}
From this angle of view, we can also see why  the scattering amplitude is not a mandatory ingredient if one's aim is only to find out discrete energy spectrum of few-hadron system.

In this notes, we aim to illustrate above mentioned three claims in a  way as simple as possible without   distractions of all the fancy and complicated technical  dress-up. In Sec.\ref{twobody}, we will start with a simple example to show how quantization condition is formulated in two-body sector in general  and it is relation to    L\"uscher  formula \cite{Luscher:1990ux}. In Sec.\ref{fewbody}, we will illustrate how the situation is complicated by subprocess  pair-wise interactions in three-body problems, and show how the quantization condition can be formulated properly and  discuss why infinite volume scattering amplitude is in fact dispensable in formulating quantization condition. A summary is given in Sec.\ref{summ}.

\section{Two-body interaction and  L\"uscher  formula}\label{twobody}

From this point on, all the discussions will be  carried out in terms of quantum mechanical operators, so that the  scattering process regardless boundary conditions  can be formulated   on an equal footing symbolically. As far as only short-range interactions are considered, {\it i.e. } interaction range is much shorter than size of box, boundary condition will only have significant impact on analytic properties of Green's functions.  Most importantly,  conclusion can be drawn based on the simple discussion even  without digging into all the distracting technical  details.

Let's first introduce a $T$-matrix operator,   scattering process is conventionally described by inhomogeneous Lippmann-Schwinger equation \cite{Newton:1982qc},
\begin{equation}
\hat{t } (E)= \hat{V} +\hat{ V} \hat{G}_0  (E) \hat{ t} (E),
\end{equation}
where $\hat{V}$ stands for potential operator.   Free Green's function  operator, $\hat{G}_0$, is given by
\begin{equation}
\hat{G}_0  (E)=  \frac{1}{E-\hat{H}_0 + i \epsilon} = \sum_\mathbf{ k}  \frac{ | \mathbf{ k} \rangle \langle \mathbf{ k} | }{E-E_\mathbf{ k} + i \epsilon}, 
\end{equation}
where $\hat{H}_0$ is free particles Hamiltonian, and $$\hat{H}_0  | \mathbf{ k} \rangle = E_\mathbf{ k} | \mathbf{ k} \rangle .$$

 In infinite volume,  eigenstates of $\hat{H}_0$, $ | \mathbf{ k} \rangle $, are continuous. Hence, infinite volume Green's function $\hat{G}^{(\infty)}_0 (E)$  develop a branch cut on real $E$ axis that ultimately split $\hat{t}^{(\infty)} (E)$  into  physical and unphysical amplitudes defined in first and second Riemann sheets respectively.  The physical amplitude is defined on real $E$ axis with $\epsilon \rightarrow 0$. Above two-particle threshold,  $\hat{G}^{(\infty)}_0 (E)$ has both continuous principle part and imaginary part,  therefore,  with a real $\hat{V}$, the solution of equation
 \begin{equation}
\frac{1}{ \hat{t }^{( \infty)} (E) }= \frac{1}{\hat{V}  } -  \hat{G}^{(\infty)}_0  (E) =0
 \end{equation}
 does not exist not only on real $E$ axis but also on complex plane in physical sheet. Hence, in infinite volume, stationary bound state solutions can be found only below two-particle threshold, where imaginary part of $\hat{G}^{(\infty)}_0 (E)$ vanishes.
 
 On the contrary, in finite volume, eigenstates   $ | \mathbf{ k} \rangle $  become  discrete, the branch cut of   $\hat{G}^{(\infty)}_0 (E)$  is replaced by discrete $\delta(E- E_\mathbf{ k})$  poles in   finite volume Green's function $\hat{G}^{(L)}_0 (E)$. The principle part of $\hat{G}^{(L)}_0 (E)$ becomes periodic, and imaginary part of $\hat{G}^{(L)}_0 (E)$ vanishes as far as $E \neq E_\mathbf{ k} $. Therefore, the solution of 
  \begin{equation}
\frac{1}{ \hat{t}^{( L)} (E) }= \frac{1}{\hat{V}  } -  \hat{G}^{(L)}_0  (E) =0 \label{2bqc}
 \end{equation}
can be   found on entire real $E$ axis   in finite volume. Equivalently, the pole position of $\hat{t }^{( L)} (E)$ are associated with stationary state solutions of homogeneous  Lippmann-Schwinger equation,
\begin{equation}
\hat{t } (E)=  \hat{ V} \hat{G}_0  (E) \hat{ T} (E). \label{homo2beq}
\end{equation}
Using relation $\hat{t } = \hat{V} \Psi$,  Eq.(\ref{homo2beq}) is thus can be converted into 
\begin{equation}
    \Psi (E)=\hat{ V}   \hat{G}_0  (E) \Psi (E), \label{homoscheq}
\end{equation}
 which describes stationary bound states of system.

 In summary, the stationary solutions can be found by either looking for pole positions of solution of inhomogeneous Lippmann-Schwinger equation,
 \begin{equation}
 \hat{t } (E) =  \frac{1}{ \frac{1}{\hat{V}  } -  \hat{G}_0  (E) },
 \end{equation}
 or equivalently solving homogeneous Lippmann-Schwinger equation, Eq.(\ref{homo2beq}) or Eq.(\ref{homoscheq}) directly. This statement in fact is true regardless boundary conditions. In infinite volume, only stationary solutions can be found are bound states below two-particle threshold. However, in finite volume, infinite discrete solutions can be found even above two-particle threshold. Therefore, finding stationary solutions is nothing but  solving    bound state problems with  certain boundary condition constraint on wave function, 
 \begin{equation}
 \hat{H} \Psi  = E   \Psi .
 \end{equation}

 The finite volume quantization condition given in Eq.(\ref{2bqc}) can be recasted  in terms of $\hat{T}^{(\infty)}$  by assuming that  short-range interaction potential $\hat{V}$ remains same in both finite and infinite volume, thus  we obtain
 \begin{equation}
  \frac{1}{\hat{V}  }  =  \frac{1}{ \hat{t }^{( \infty)} (E) } + \hat{G}^{(\infty)}_0  (E). 
 \end{equation}
 The  finite volume quantization condition now is given by
   \begin{equation}
  \frac{1}{ \hat{t }^{( \infty)} (E) } + \hat{G}^{(\infty)}_0  (E) -  \hat{G}^{(L)}_0  (E) =0, \label{luscher}
 \end{equation}
 which is nothing but  L\"uscher  formula \cite{Luscher:1990ux}. In two-body case, due to the fact that the interaction potential and $\hat{t}$ has a relatively simple relation, so the quantization condition can be formulated in terms of infinite volume scattering amplitude  $\hat{t }^{( \infty)}$ directly.

\section{Few-body interaction}\label{fewbody}  

\subsection{$T$-matrix formalism}
It is very tempting to extend and generalize the previous described procedure into few-body sectors, unfortunately the situation in few-body case is complicated by   physical processes  where  some particles are disconnected from and not interacting with rest of particles. The conclusion and argument we are going to draw and make in fact can be made in general, however, for the sake of simplicity, we will only use three-body interaction as a simple example in follows. It is sufficient to just make a  point.

Let's  now consider three-particle interacting with pair-wise interactions, 
three-particle scattering process again is described by inhomogeneous Lippmann-Schwinger equation \cite{9780706505740,Faddeev:1960su,Newton:1982qc},
\begin{equation}
\hat{T } (E)= \hat{V} +\hat{ V} \hat{G}_0  (E) \hat{ T} (E), \label{3bLippmann}
\end{equation}
where $$\hat{V} = \sum_{i=1}^3 \hat{V}_i   $$ and $\hat{V}_i $ stands for interaction potential operator between j-th and k-th particles.    It has been a well-known fact that pair-wise potentials yield the appearance of a $\delta$-function in the kernel of Eq.(\ref{3bLippmann})  because of   momentum conservation of  third spectator particle. This $\delta$-function persist during all orders of iterations,  hence, the kernel $\hat{ V} \hat{G}_0  (E)$ in three-body case is not compact, and Lippmann-Schwinger equation in three-body case cannot be solved by Fredholm method \cite{9780706505740,Faddeev:1960su,Newton:1982qc}.   To overcome this difficulty, Faddeev approach normally is introduced by splitting  Eq.(\ref{3bLippmann}) into coupled equations,
\begin{equation}
\hat{T }_i (E)= \hat{V}_i +\hat{ V}_i \hat{G}_0  (E) \hat{ T} (E),  \ \ \ \  i = 1,2,3,
\end{equation}
and  $$ \hat{ T} (E) = \sum_{i=1}^3 \hat{T }_i (E) .$$
Rearranging  above equations, one    obtain inhomogeneous Faddeev equations  
\begin{align}
& \begin{bmatrix}
\hat{T }_1 (E) \\
\hat{T }_2 (E) \\
\hat{T }_3 (E)
\end{bmatrix}
=\begin{bmatrix}
\hat{t }_1 (E) \\
\hat{t }_2 (E) \\
\hat{t }_3 (E)
\end{bmatrix} 
+ 
\hat{ \mathcal{K} }(E)   \hat{G}_0  (E)
\begin{bmatrix}
\hat{T }_1 (E) \\
\hat{T }_2 (E) \\
\hat{T }_3 (E)
\end{bmatrix},
\end{align}
now with a compact kernel matrix  
\begin{align}
& \hat{ \mathcal{K} }(E) =
\begin{bmatrix}
 0 &  \hat{t }_1 (E)&\hat{t }_1 (E) \\
\hat{t }_2 (E) & 0 & \hat{t }_2 (E) \\
\hat{t }_3 (E) & \hat{t }_3 (E) & 0
\end{bmatrix}  , \label{kernel}
\end{align}
where $ \hat{t }_i (E)$  is subprocess amplitude and  given by $$\hat{t }_i (E) = \frac{1}{ \frac{1}{ \hat{V}_i}  - \hat{G}_0  (E) }.$$
    The solutions of $\hat{T}$-matrix   is   hence given by
\begin{align}
& \begin{bmatrix}
\hat{T }_1 (E) \\
\hat{T }_2 (E) \\
\hat{T }_3 (E)
\end{bmatrix}
= \left [  \mathbb{I} -\hat{ \mathcal{K} }(E)    \hat{G}_0  (E)  \right ]^{-1}  \begin{bmatrix}
\hat{t }_1 (E) \\
\hat{t }_2 (E) \\
\hat{t }_3 (E)
\end{bmatrix} . \label{faddeeveq}
\end{align}
The stationary state solutions   are  associated with the pole positions of Faddeev equations solutions, hence are determined by finding solutions of
\begin{equation}
\det \left [ \mathbb{I} -\hat{ \mathcal{K} }(E) \hat{G}_0  (E)  \right ]=0.
\end{equation} 
Similar to tow-body case, in infinite volume, stationary state solutions can only be found below three-body threshold as bound state solutions. In finite volume, stationary state solutions  exist even above   three-body threshold due to periodic boundary condition.
Equivalently,  stationary state solutions can also be found by using homogeneous Faddeev equations directly,
\begin{align}
& \left [  \mathbb{I} -\hat{ \mathcal{K} }(E)    \hat{G}_0  (E)  \right ]
\begin{bmatrix}
\hat{T }_1 (E) \\
\hat{T }_2 (E) \\
\hat{T }_3 (E)
\end{bmatrix} =0. 
\end{align}

Although three-body quantization condition 
\begin{equation}
\det \left [  \frac{1}{ \hat{\mathcal{K}} (E) } - \hat{G}_0  (E)  \right ]=0, 
\end{equation} 
resemble two-body quantization condition given in Eq.(\ref{2bqc}), unlike two-body case, $\hat{\mathcal{K}}  $ now depend on $ \hat{t }_i $, see Eq.(\ref{kernel}).  Thus, $\hat{\mathcal{K}}  $ is   affected by boundary condition as well,   infinite volume $\hat{\mathcal{K}}^{(\infty)}  $ and finite volume $\hat{\mathcal{K}}^{(L)}  $   have quite different analytical properties.  Therefore, a simple  L\"uscher  formula type quantization condition is no longer available in few-body case. The few-body quantization condition cannot be easily  formulated in terms of infinite volume scattering amplitude $\hat{ T}^{(\infty)} $ directly.  However, the connection between finite and infinite volume dynamics can be made again  based on the assumption that potentials  remain short-range and same  in both finite volume and infinite volume. Hence,  subprocess amplitudes $\hat{t }^{(L)}_i  $ can be related to $\hat{t }^{(\infty)}_i  $ by
\begin{equation}
\frac{1}{\hat{t }^{(L)}_i (E) }=   \frac{1}{\hat{t }^{(\infty)}_i (E) }  +  \hat{G}^{(\infty)}_0  (E)  - \hat{G}^{(L)}_0  (E) .
\end{equation}
The   $\hat{\mathcal{K}}^{(L)} $  now can be parameterized by using either $\hat{V}_i$ or $\hat{t }^{(\infty)}_i  $ as basic ingredients,  and the finite volume  quantization condition   can be formulated in terms of either $\hat{V}_i$ or $\hat{t }^{(\infty)}_i  $ as well. After the determination of basic ingredients,  $\hat{V}_i$ or $\hat{t }^{(\infty)}_i  $ by fitting lattice results, if one is interested in obtaining  the infinite volume three-body scattering amplitude as well, then  $\hat{T}^{(\infty)}  $ has to be computed in a separate step by using infinite volume inhomogeneous Faddeev equations, Eq.(\ref{faddeeveq}).

In summary, infinite volume few-body scattering  amplitude is in fact not a necessary component when it comes to the formulating  few-body quantization condition. The finite volume few-body quantization condition
$$ \det \left [  \frac{1}{ \hat{\mathcal{K}}^{(L)} (E) } - \hat{G}^{(L)}_0  (E)  \right ]=0 $$
 can be obtained in terms of  some basic ingredients of  infinite volume few-body scattering amplitudes alone. The  infinite volume few-body scattering amplitude can be computed separately by Faddeev equations after basic ingredients are determined. These basic ingredients can be chosen either as interaction potential $\hat{V}_i$ or their associated subprocess amplitudes  $$\hat{t }^{(\infty)}_i  (E) =  \left [  \hat{V}_i^{-1}  - \hat{G}^{(\infty)}_0 (E) \right ]^{-1}. $$

\subsection{$K$-matrix formalism}
The three-body scattering process may also be described by using $K$-matrix formalism \cite{Kowalski:1972dj,10.1143/PTP.34.284}, 
\begin{equation}
\hat{K } (E)= \hat{V} +\hat{ V} \hat{G}_{0 \mathcal{P} }  (E) \hat{ K} (E),
\end{equation}
where $\hat{G}_{0 \mathcal{P} }  (E)$ stands for the principle part of $ \hat{G}_0  (E)$, and $\hat{T}$-matrix and $\hat{K}$-matrix are related by
\begin{equation}
\hat{T } (E)=   \hat{ K} (E)  - i \pi \delta (E- \hat{H}_0)   \hat{ K} (E)  \hat{ T} (E).
\end{equation}
After carrying out Faddeev's procedure, $\hat{K } = \sum_{i=1}^3 \hat{K }_i$,  the solution of $\hat{K}$-matrix is given by
\begin{align}
& \begin{bmatrix}
\hat{K }_1 (E) \\
\hat{K }_2 (E) \\
\hat{K }_3 (E)
\end{bmatrix}
= \left [ \mathbb{I} -\hat{ \mathcal{K} }_\mathcal{P} (E)    \hat{G}_{0 \mathcal{P} }  (E) \right ]^{-1}  \begin{bmatrix}
\hat{t }_{1\mathcal{P} } (E) \\
\hat{t }_{2 \mathcal{P} } (E) \\
\hat{t }_{3 \mathcal{P} } (E)
\end{bmatrix} ,
\end{align}
where $$\hat{t }_{i  \mathcal{P}}  (E) = \frac{1}{ \frac{1}{\hat{V}_i } - \hat{G}_{0 \mathcal{P}} (E) } $$ and 
\begin{align}
& \hat{ \mathcal{K} }_\mathcal{P}(E) =
\begin{bmatrix}
 0 &  \hat{t }_{1\mathcal{P}}  (E)&\hat{t }_{1 \mathcal{P}} (E) \\
\hat{t }_{2 \mathcal{P}} (E) & 0 & \hat{t }_{ 2 \mathcal{P}} (E) \\
\hat{t }_{3 \mathcal{P}} (E) & \hat{t }_{3 \mathcal{P}} (E) & 0
\end{bmatrix}  .
\end{align}
Hence, we can conclude that $K$-matrix formalism is equivalent to $T$-matrix formalism and doesn't change the picture fundamentally. In the end, the quantization condition can be formulated in terms of either  $\hat{V}_i$ or $\hat{t }^{(\infty)}_{i \mathcal{P}}    $ as well, and infinite volume scattering amplitude still need to be computed in a separate step.

\section{Summary}\label{summ}  

In summary, we have illustrated that except in two-body  case,  infinite volume few-body scattering amplitudes are  in fact not  a mandatory component for formulating  finite volume quantization condition.  The essential ingredients of formulating finite volume quantization condition can be chosen as either interactions potential $\hat{V}_i $ or corresponding subprocess amplitudes  $$\hat{t}_i^{(\infty)} = \left  [ \hat{V}^{-1}_i - \hat{G}_0 \right ] ^{-1}.$$ The total few-body amplitude $\hat{T}^{(\infty)}$  can be computed in a separate step. In fact, if only obtaining quantization condition is one's aim,    few-body amplitude $\hat{T}^{(\infty)}$  is not needed at all.  It comes as no surprise due to the fact that   the final few-body physical process is in fact generated by all subprocess or interactions.  In the end, it all comes down to the problem of finding stationary solutions of system with periodic boundary condition constaint, which can be accomplished by either looking for the pole of solutions of inhomogeneous Faddeev equations or solving homogeneous equations directly.

\bibliography{ALL-REF.bib}

\end{document}